\documentclass[twocolumn,english,aps,prl,superscriptaddress]{revtex4}
\usepackage{amsmath}
\usepackage{graphicx}
\usepackage{amssymb}
\usepackage{textcomp}
\usepackage{esint}
\usepackage{babel}


\begin{document}

\title{Thermalization dynamics close to a quantum phase transition}
\author{Dario Patan\`e}
\affiliation{MATIS CNR-INFM $\&$ Dipartimento di Metodologie Fisiche e Chimiche (DMFCI),
Universit\`a di Catania, viale A. Doria 6, $I-95125$ Catania, Italy}
\affiliation{Departamento de F\'isica de Materiales, Universitad Complutense de Madrid, $%
E-28040$ Madrid, Spain}
\author{Alessandro Silva}
\affiliation{The Abdus Salam International Centre for Theoretical Physics, Strada
  Costiera $11$, $I-34100$ Trieste, Italy }
\author{Fernando Sols}
\affiliation{Departamento de F\'isica de Materiales, Universitad Complutense de Madrid, $%
E-28040$ Madrid, Spain}
\author{Luigi Amico}
\affiliation{MATIS CNR-INFM $\&$ Dipartimento di Metodologie Fisiche e Chimiche (DMFCI),
Universit\`a di Catania, viale A. Doria 6, $I-95125$ Catania, Italy}
\affiliation{Departamento de F\'isica de Materiales, Universitad Complutense de Madrid, $%
E-28040$ Madrid, Spain}
\begin{abstract}
We investigate the dissipative dynamics of a quantum critical system in
contact with a thermal bath. In analogy with the standard protocol employed to analyze aging, we study the response of a system to a sudden change of the bath temperature.
The specific example of the XY model in a transverse magnetic field whose spins are locally coupled to a set of bosonic baths is considered. 
The peculiar nature of the dynamics is encoded in the correlations developing out of equilibrium. By means of a kinetic equation we analyze the spin-spin correlations and block correlations. 
We identify some universal features in the out-of-equilibrium dynamics. Two distinct regimes, characterized by different time and length scales, emerge. During the initial transient the dynamics is characterized by the same critical exponents as those  of the equilibrium quantum phase transition and resembles the dynamics of thermal phase transitions. At long times equilibrium is reached through the propagation along the chain of a thermal front in a manner similar to the classical Glauber dynamics.
\end{abstract}

\maketitle

\emph{Introduction.} Understanding the out-of-equilibrium dynamics
of quantum many-body systems is a central issue of modern condensed
matter physics from both a fundamental and an applicative point
of view. Theoretical interest on these problems traces back to the
studies of irreversibility in non equilibrium thermodynamics.
In quantum systems the interplay between phase coherence, 
strong interactions, and low dimensionality may result in surprising 
dynamical behaviors \cite{NONthermalization}.
 Remarkably, this kind of issues can be explored
experimentally at the quantum level by realizing highly controllable
quantum many-particle systems. In this sense, cold atoms in
optical lattices are the paradigmatic example of an interacting
system where the interaction strength and 
the geometrical  settings can be fine tuned. 
The engineered Hamiltonians can mimic condensed matter systems
\cite{Lewenstein07} and also provide feasible tools to investigate 
many interesting issues in non-equilibrium statistical mechanics.
\cite{coldatoms}.
 Similarly, arrays of coupled microcavities have been shown to have the potential to act
as simulators of quantum many-body dynamics, with characteristics
complementary to those of optical lattices \cite{Hartmann06}.

In this context it is desirable to consider simple but illustrative enough situations. 
An important
problem that has been studied is the response of a system to a
\emph{strong} perturbation  driving it out of equilibrium.
Paradigmatically, this occurs when a given system is forced out of equilibrium by a sudden change of a control parameter ({\it quantum quench}) \cite{quantum_quenches}. The richness and complexity of
this problem is ultimately related to the nonlocal character of the
correlations developing the system temporal evolution. An elegant  picture of the phenomenon  
has been provided  by Calabrese and Cardy \cite{Calabrese07},  according 
to which  correlations among spatially distant parts of the system are established because, from each point, pairs of entangled quasiparticles are 
emitted and 
propagate ballistically in opposite directions  with a 
characteristic velocity.
This  picture, which allows to
describe the dynamics of correlation functions and block entropy,
has been tested in different contexts (see \cite{Calabrese08} and
references therein).

The above results concern the idealized situation of the unitary
dynamics of a closed system. 
A central problem then, is to  understand   the influence of an
external environment on the dynamics. This issue, originally
explored in the context of mean-field spin glasses
\cite{Cugliandolo98,Kennett01}, has very recently obtained some
attention in the context of the adiabatic dynamics of open quantum
critical systems \cite{Fubini07,Mostame07,Amin08,PatanePRL,Cincio08}.

In this Letter we study the problem of a \emph{{}thermal quench}, i.e. the
response of the system to a sudden change of the
temperature keeping the Hamiltonian parameters fixed. This
is the purely dissipative analog of a quantum quench \cite%
{thermalquench}. When the temperature of the bath in
contact with the system is changed, the system is forced to relax
towards the new thermal equilibrium state. We will be focusing on such a thermalization process close to a quantum phase transition.  Some key questions we wish to address are: How is thermalization achieved
microscopically? What are the characteristic time and length scales
emerging from the out-of equilibrium dynamics? And, ultimately: What are the
dynamical signatures of the quantum phase transition?

We address the questions above for a benchmark class of systems with
a quantum critical point, namely, the quantum XY model. To model a
thermal reservoir the system is coupled to a bosonic bath. As in the
unitary case (conservative system), we will focus on the study of spin-spin correlation
functions. A
detailed study of the dynamics of the system will allow us to understand
how not only quasi-equilibrium but also far-from-equilibrium
dynamics reveals signatures of criticality.

\emph{Model.} We consider a chain of $N$ spins with an anisotropic Ising
interaction and subject to a transverse magnetic field:
\begin{equation}
H_{S}=-\frac{\mathcal{J}}{2}\!\sum_{j}^{N}\!\left(\frac{1+\gamma}{2}%
\sigma_{j}^{x}\sigma_{j+1}^{x}+\frac{1-\gamma}{2}\sigma_{j}^{y}%
\sigma_{j+1}^{y}\!+\! h\sigma_{j}^{z}\right) \,  \label{eq: H_S}
\end{equation}
where $\sigma^{x,y,z}$ are Pauli matrices. We fix the energy scale $\mathcal{%
J}=1$ and consider $h>0$. In the anisotropic case $0<\gamma\leq1$, the
magnetic field $h$ induces a phase transition at $h_{c}=1$ that separates a
paramagnetic phase at $h>1$ from a ferromagnetic ordered phase with $%
\left\langle \sigma^{x}\right\rangle \ne0$; such a phase transition belongs
to the Ising universality class with critical indexes $\nu=z=1$. The
Hamiltonian can be diagonalized in momentum space, in terms of Jordan-Wigner
fermions $c_{k}$, as $\sum_{k>0}\Psi_{k}^{\dagger}\mathcal{\hat{H}}%
_{k}\Psi_{k}$, where $\Psi_{k}^{\dagger}=\left(c_{k}^{\dagger},\
c_{-k}\right)$, $\mathcal{\hat{H}}_{k}=-(\cos k+h)\hat{\tau}^{z}+\sin k\
\hat{\tau}^{y}$, where $\hat{\tau}^{\alpha}$ are Pauli matrices. The $T=0$
quantum phase transition leaves an imprint at low temperatures, leading,
close to the quantum critical point, to a crossover at temperatures $%
T\sim\Delta$ with $\Delta \equiv |h-h_{c}|$ the energy gap. 
In particular for $T\ll\Delta$ the spin-spin correlation function is factorized
into a quantum and a thermal term that can be described semiclassically in terms of
quasiparticle excitations, while in the quantum critical region ($T\gg\Delta$)
 quasiparticle excitations no longer exist~\cite{SachdevBOOK}.

To model a thermal reservoir we consider a set of bosonic baths coupled
locally to each spin \cite{PatanePRL}, such that the global Hamiltonian
reads
\begin{equation}
H=H_{S}+\sum_{j}^{N}\! X_{j}\sigma_{j}^{z}+H_{B}\;.
\label{eq:HamiltonianKap}
\end{equation}
where $X_{j}=\sum_{\beta}\lambda_{\beta}(b_{\beta,j}^{\dagger}+b_{\beta,j})$
and $H_{B}=\!\sum_{j,\beta}\!\omega_{\beta}b_{\beta j}^{\dagger}b_{\beta j}$%
. The system-bath coupling is chosen to have power law spectral densities $%
\sum_{\beta}\lambda_{\beta}^{2}\delta(\omega-\omega_{\beta})=2\alpha%
\omega^{s}\exp(-\omega/\omega_{c})$ \cite{WeissBOOK}. The
system-bath coupling
we are considering, Eq.~(\ref%
{eq:HamiltonianKap}), breaks the integrability of the model,
inducing transitions between all energy levels, and thus complete
relaxation.

The quantum quench dynamics for the closed XY model was studied in \cite%
{quenchXY}. It is customary to consider
the physical system initially uncorrelated, e.g. by applying a strong
magnetic field $h$. After a quench of the magnetic field, correlations
between parts of the system will start to develop because of the dynamics
induced by the new Hamiltonian. Analogously, in the case of thermal
quenches, we consider the system to be initially prepared in equilibrium
with the bath at a very high temperature, again with no correlations because
the density matrix of the system is proportional to the identity. After a quench of the
temperature of the bath at $t=0$, the system is forced out of equilibrium
and eventually reaches a new stationary thermal state. In the following we
investigate how such correlations develop and how thermal equilibrium is
eventually approached. All the results shown in the figures below refer to
the Ising model ($\gamma =1$) coupled to Ohmic baths ($s=1$). However, the
results stated in the text refer to the general case $0<\gamma \leq 1$, $s>0$%
. We discuss the spin-spin correlation function and later we consider the quantum mutual information.

\begin{figure}[tbp]
\includegraphics[scale=0.7]{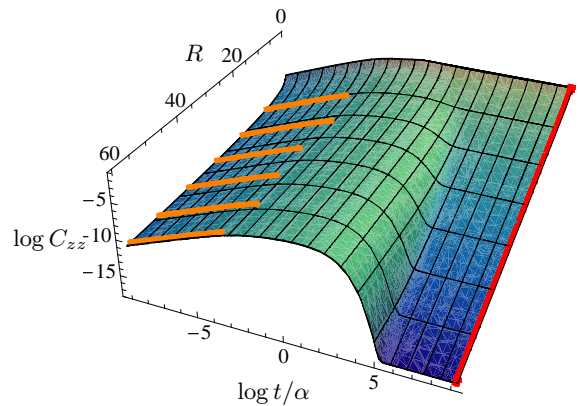}\put(-130,2){$\log t/\alpha$}
\put(-215,60){$\log C_{zz}$} \put(-150,130){$R$}
\caption{Correlation function $C_{zz}$ for a quench from $T=\infty$ to $%
T=0.1 $ at $h=1$; red thick line on the right at large $t$ is the thermal
equilibrium $C_{zz}(R)$. During the initial transient ($t\ll\protect\alpha$%
), $C_{zz}\propto t^{2}$ for all $R$, as marked by the fits on the left. }
\label{fig:capitalistic1}
\end{figure}

\emph{Spin-spin correlations.} We consider the equal-time connected
correlation function%
\begin{equation}
C_{zz}(t,\ R)=\left\langle
\sigma_{j}^{z}(t)\sigma_{j+R}^{z}(t)\right\rangle -\left\langle
\sigma_{j}^{z}(t)\right\rangle \left\langle
\sigma_{j+R}^{z}(t)\right\rangle\,.  \label{eq:ZZc}
\end{equation}
In the case of thermal quenches the dynamics is purely dissipative.
Since for weak coupling $\alpha\ll1$ the dynamics of populations and
coherences decouple, if the system starts in a mixed state no
coherences will develop after the quench (this is consistent with
the so called ``secular approximation'' \cite{CohenBOOK}).
Therefore in this limit at each time the system is approximately in
a statistical mixture of the Hamiltonian eigenvectors, i.e. a
gaussian state. Hence, by exploiting this the correlation function can
be expressed as $%
C_{zz}=\left| \frac{4}{N}\sum_{k>0}\langle c_{k}c_{-k}\rangle
\sin(kR)\right|^2-\left(\frac{4}{N}\sum_{k>0}\cos(kR) \langle
c_{k}^{\dagger}c_{k}\rangle\right)^2$. In order to evaluate the two
point fermionic correlators we use the kinetic equation derived in \cite%
{PatanePRL,PatanePRL_long} within the weak coupling and Markov
approximations. From the analysis of our results (Fig.~\ref%
{fig:capitalistic1}) two regimes can be outlined: right after the quench,
for $t\ll\alpha$, correlations increase as $C_{zz}\propto t^{2}$, while in
the opposite limit, at times $t\gg\alpha$, the system is close to thermal
equilibrium. During the initial transient, $C_{zz}$ reaches for far distant
spins ($R\gg1$) values greater than those of thermal equilibrium, $%
C_{zz}(R)>C_{zz}^{\mathrm{th}}(R)\propto\exp(-R/\xi)$, with $\xi$
the thermal correlation length \cite{SachdevBOOK}. Thus, the
crossover to the long-time regime displays a non-monotonous behavior
as a function of time, so that that $C_{zz}$ increases up to a
maximum value and then relaxes to the thermal equilibrium value
(Fig.~\ref{fig:capitalistic1}).

\begin{figure}[tbp]
\hspace{-0.2cm}\includegraphics[scale=0.5]{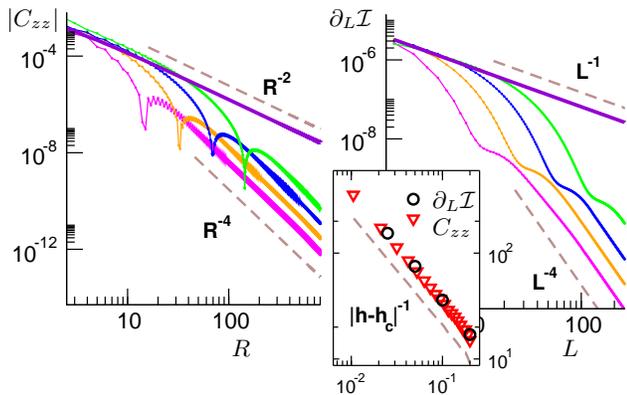}\put(-235,140){$%
|C_{zz}|$}\put(-113,140){$\partial_{L}\mathcal{I}$}\put(-74,71){{\small $%
\partial_{L}\mathcal{I}$}}\put(-74,61){{\small $C_{zz}$}}\put(-150,15){$R$%
}\put(-25,15){$L$}
\caption{Initial transient: snapshots of $C_{zz}$ (left) and $\partial_{L}%
\mathcal{I}$ (right) at a fixed $t/\protect\alpha\ll1$ after a quench from $%
T=\infty$ to $T=0.1$ and for $h=0.8,\ 0.9,\ 0.95,\ 0.975,\ 1$ (from
bottom to top). The spikes relative to $C_{zz}$ in the left panel mark the
distances $\protect\xi_{t}(h)$ at which $C_{zz}$ changes sign. Dashed lines
are plotted for comparison. \emph{Inset:} $\protect\xi_{t}$ as a function of
$|h-h_{c}|$; for $\partial_{L}\mathcal{I}$, $\protect\xi_{t}$ is calculated
as the maximum of $\partial_{L}^{2}\mathcal{I}$ which marks the crossover
between $L^{-1}$ and $L^{-4}$ scaling.}
\label{fig:Capitalistic_transient}
\end{figure}

Let us analyze first the initial transient. We observe that, for noncritical
values of the magnetic field ($h\ne1$), $C_{zz}$ changes its sign at a
certain distance $\xi_{t}(h)$ such that $C_{zz}\lessgtr0$ for $%
R\gtrless\xi_{t}$ (see Fig. \ref{fig:Capitalistic_transient}). That distance
marks the crossover between two power-law behaviors with different exponent,
respectively $R^{-4}$ and $R^{-2}$, and close to the critical point it
diverges as
\begin{equation}
\xi_{t}\propto|h-h_{c}|^{-1}.  \label{eq:xi_t}
\end{equation}
Collecting all the above results, we find that the long $R$ behavior close
to the critical point is described by
\begin{equation}
C_{zz}\propto t^{2}%
\begin{cases}
R^{-2} & \ R\ll\xi_{t} \\
R^{-4} & \ R\gg\xi_{t}%
\end{cases}%
\ \ \   \label{eq:Czz transient FINAL}
\end{equation}
At equilibrium, close to the phase transition, the correlation
length $\xi$ marks the crossover between the exponential decay for
$R\gg\xi$ and the critical power-law for $R\ll\xi$. Similarly, in the
present non-equilibrium case $\xi_{t}$ can be interpreted as an
effective crossover scale between two power-law regimes. 
Eqs. (\ref{eq:xi_t}) and (\ref{eq:Czz transient FINAL}) are
independent of the specific value of the final temperature at which
the system is quenched and of the specific exponent $s$ of the bath
spectral function. Moreover, they are robust within the range
$0<\gamma\leq1$ in which the system belongs to the Ising
universality class. 
 Remarkably, although in this regime the system is far from equilibrium,
Eqs. (\ref{eq:xi_t}) and (\ref{eq:Czz transient FINAL}) are characterized by 
the equilibrium critical indexes: $\xi\propto |h-h_c|^{-1}$ and $C_{zz}\propto R^{-2},\ R\ll\xi$.
\newline

\begin{figure}[tbp]
\includegraphics[scale=0.48]{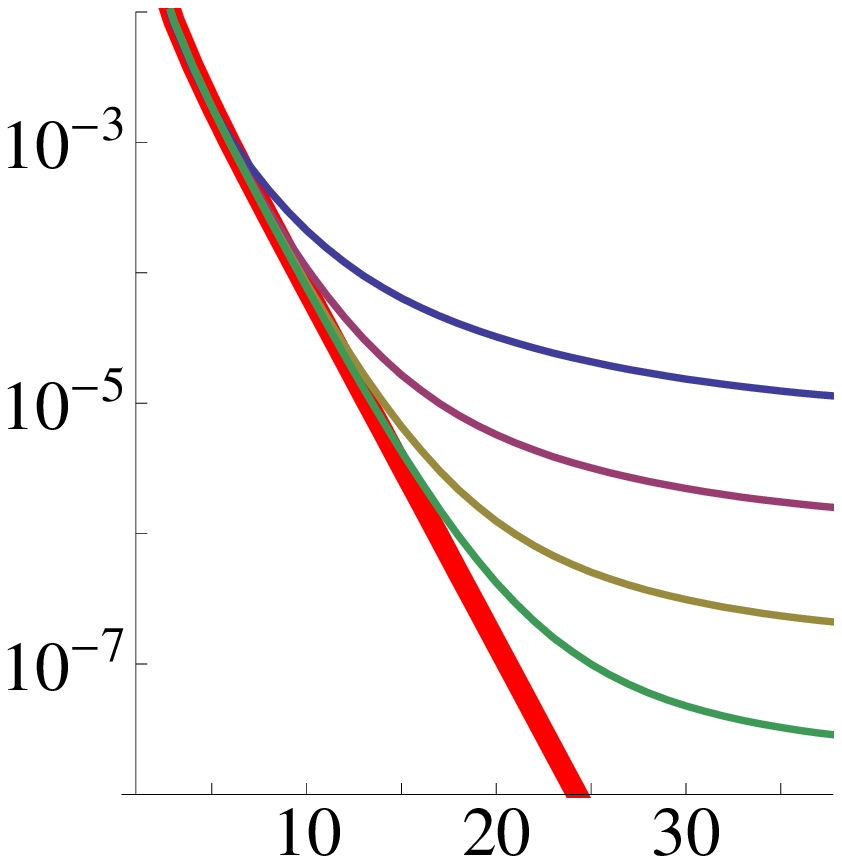}\put(-120,115){$%
C_{zz}$}\put(-30,-10){$R$}\put(-20,45){{\Huge $\downarrow$}}\hspace{0.4cm}%
\includegraphics[scale=0.48]{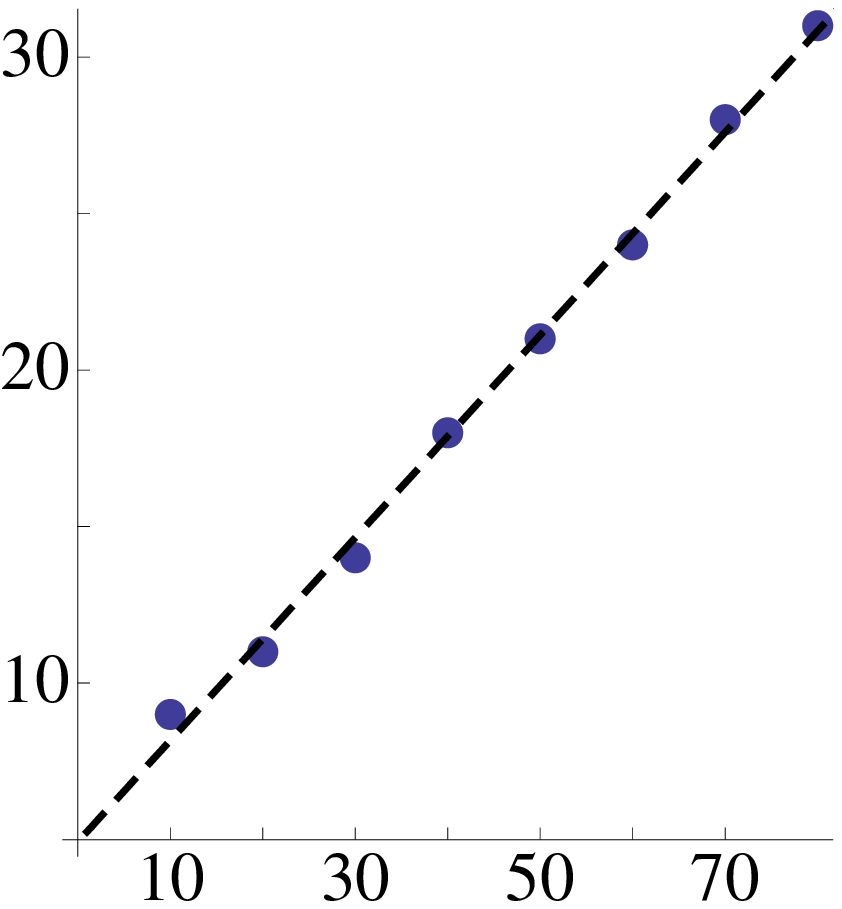}\put(-125,105){$R_{{\rm th}}$%
}\put(-30,-10){$t/\alpha$}
\caption{Thermalization of the spin-spin correlation function $C_{zz}$ after
a quench from $T=\infty$ to $T=0.1$ at $h=1$. Left: snapshots of $C_{zz}(R)$
at $t/\protect\alpha=10,\ 20,\ 30,\ 40$ from top to bottom; thick red line
is the (exponential) thermal equilibrium $C_{zz}^{\mathrm{th}}$. At a given
time after the quench $C_{zz}$ is thermalized up to a distance $R_{\mathrm{th%
}}$ that increases with time. Right: corresponding time dependence of $R_{%
\mathrm{th}}$; the linear fit gives $v_{\mathrm{th}}\simeq0.32$. }
\label{fig:Capitalistic_thermalizationfront}
\end{figure}

\begin{figure}[tbp]
\includegraphics[scale=0.42]{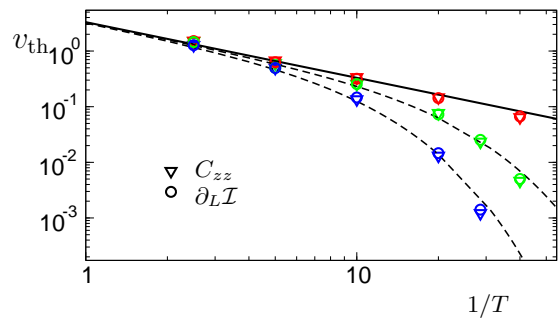}\put(-207,90){{\large $v_{{\rm th}}$%
}}\put(-35,-10){$1/T$}\put(-138,41){$C_{zz}$}\put(-138,31){$\partial_{L}%
\mathcal{I}$}
\caption{Thermal front velocity $v_{\mathrm{th}}$ as a function of $1/T$.
From top to bottom $h=1,\ 0.9,\ 1.2$ (so that $\Delta=0,\ 0.1,\ 0.2$). Lines
are the fit $\ Ta(1+b\frac{\Delta}{T})\ e^{-\Delta/T}$ with $a=3.3,\ b=0.9$
for the specific case $\protect\gamma=1,\ s=1$.}
\label{fig:Capitalistic_vth}
\end{figure}

We now focus on the long time regime. The analysis of Fig. \ref%
{fig:Capitalistic_thermalizationfront} indicates that, at a given time, the
correlation function is thermalized up to a distance $R_{\mathrm{th}}$. This
\emph{thermal front} exists because the long-distance correlations are
dominated by the slowly relaxing low-energy modes. The front is found to
propagate  ballistically
with a speed $v_{\mathrm{th}}$ that is a function of $T$
and $h$. In particular, as shown in Fig. \ref{fig:Capitalistic_vth}, the
velocity scales as
\begin{equation}
v_{\mathrm{th}}\propto%
\begin{cases}
T^{s} & T\gg\Delta \\
e^{-\Delta/T} & T\ll\Delta%
\end{cases}
\label{eq: capitalistic vth}
\end{equation}
where $\Delta=|h-1|$ is the energy gap of the XY model.

\emph{Block correlations.} We now analyse how correlations between a block
of spins and the rest of the chain develop after a thermal quench. In order
to quantify such a correlation, we use a tool originally developed in the
context of quantum information theory. For a certain bipartition of the
system into two blocks of $L$ and $N-L$ spins, the mutual information is
defined as%
\begin{equation}
\mathcal{I}(L)=S(\rho_{L})+S(\rho_{N-L})-S(\rho_{N})  \label{eq:mutual}\; ,
\end{equation}
where $S(\rho)=-\mathrm{Tr}\left( \rho\log\rho\right) $ and $\rho_{N}$ is
the density matrix of the entire system. 
The mutual information measures the correlations between the two blocks of $L$ and $N-L$
spins \cite{Wolf08}. For the XY model $\mathcal{I}(L)$ 
 is known to diverge logarithmically as a function of $L$ at the critical
point, while it saturates for noncritical values. In the following we
concentrate on the derivative $\partial_{L}\mathcal{I}$, which measures the
sensitivity of the correlations to the block size. It is useful to study $%
\partial_{L}\mathcal{I}$ because, at equilibrium, it shows features similar
to the spin-spin correlation function: it scales as $\partial_{L}\mathcal{I}%
\propto L^{-1}$ at the critical point, 
while it decays exponentially for noncritical values.

Equation (\ref{eq:mutual}) can be computed in terms of two-point fermionic
correlators (obtained by solving the kinetic equation) using the results of
Ref. \cite{Peschel03}. We study for the mutual information the same setting
of thermal quenches investigated above for the spin-spin correlation
function. Remarkably the scenario emerging is very similar to that depicted
in Fig. \ref{fig:capitalistic1}. There are two regimes: an initial transient
governed by
\begin{equation}
\partial_{L}\mathcal{I}\propto t^{2}%
\begin{cases}
L^{-1} & \ L\ll\xi_{t} \\
L^{-4} & \ L\gg\xi_{t}%
\end{cases}%
\ \ \   \label{eq:I transient FINAL}
\end{equation}
with the same correlation length (\ref{eq:xi_t}) (see Fig. \ref%
{fig:Capitalistic_transient}). In the quasi-equilibrium regime at long
times, $\partial_{L}\mathcal{I}$ exhibits a thermal front propagation
similar to that shown in Fig. \ref{fig:Capitalistic_thermalizationfront}
and with the same velocity found for $C_{zz}$ (see Fig. \ref%
{fig:Capitalistic_vth}).

\emph{Conclusions.} We have analyzed the dynamics of
spin-spin and block correlation functions following a sudden cooling
of the bath coupled to a quantum system.  
For both quantities we find that the
dynamics displays two regimes: at short times
the correlations develop
according to (\ref{eq:Czz transient FINAL}) and (%
\ref{eq:I transient FINAL}), while at long times  a
well defined thermal front propagates along the chain with 
velocity (\ref{eq: capitalistic vth}), the latter being 
sensitive to  the critical properties of the system. 
 We remark that  the system does not exhibit aging because it is quenched 
away from the critical point.  Nevertheless in its early stages relaxation 
does show  critical features analogously to those of thermal phase transitions.
In particular, for systems quenched at the critical temperature, 
it is known that equal-time two-point correlation
function scales, during the initial transient, as a power law both in time $%
\propto t^{a}$ (here $a=2$) and in space $R^{-d+2-\eta}$ (in our case $%
R^{-2} $)
\cite{thermalquench}.
 Besides, we point out that the scaling of the thermalization velocity
$v_{\mathrm{th}}\propto\exp(-\Delta/T)$, which we find in the semiclassical regions $T\ll\Delta$, holds also in
the classical Ising model within the Glauber dynamics \cite{Glauber}.
This similarity can be ascribed to the fact that the
system-bath coupling generates incoherent relaxation without
conserving the order parameter, as happens in the phenomenological Glauber
model for the dynamics of the classical spins.

\emph{Acknowledgments.} We thank R. Fazio, G.E. Santoro and P. Calabrese for discussions and comments.
D.P. acknowledges the ESF (INSTANS) for financial support. L.A. and F.S.
acknowledge support from MEC (FIS2007-65723).

\end{document}